\documentclass[preprint]{aastex} 
\begin{document}

\title{A Comparative Study of the Absolute-Magnitude Distributions of Supernovae}

\author{Dean Richardson, David Branch, Darrin Casebeer, Jennifer
Millard, R. C. Thomas and E.~Baron}

\affil{Dept. of Physics and Astronomy, University of Oklahoma,
Norman, OK 73019; richards@nhn.ou.edu}

\begin{abstract} The Asiago Supernova Catalog is used to carry out a 
comparative study of supernova absolute-magnitude distributions. An
overview of the absolute magnitudes of the supernovae in the current
observational sample is presented, and the evidence for subluminous
and overluminous events is examined. The fraction of supernovae that
are underluminous ($M_B > -15$) appears to be higher (perhaps much
higher) than one fifth but it remains very uncertain.  The fraction
that are overluminous ($M_B < -20$) is lower (probably much lower)
than 0.01.  The absolute-magnitude distributions for each supernova
type, restricted to events
within 1 Gpc, are compared. Although these distributions are
affected by observational bias in favor of the more luminous events,
they are useful for comparative studies.  We find mean absolute blue
magnitudes (for $H_0=60$) of $-19.46$ for normal Type~Ia 
supernovae (SNe~Ia), $-18.04$ for SNe~Ibc, $-17.61$ and $-20.26$ for 
normal and bright SNe~Ibc considered separately,  
$-18.03$ for SNe~II-L, $-17.56$ and $-19.27$ for normal and bright
SNe~II-L considered separately, $-17.00$ for SNe~II-P, and $-19.15$ for
SNe~IIn.
\end{abstract}


\section{Introduction}

The absolute-magnitude distributions of the various supernova (SN) types
provide vital information for determining SN rates, for
advancing our knowledge of the stellar progenitors and their explosion
mechanisms, and for planning future ground- and space-based SN 
searches. More than a decade ago, \citet[MB90;][]{mb90} used the 
Asiago Supernova Catalog \citep[the ASC;][]{barbon89} to carry out a
comparative study of absolute-magnitude distributions. At the end of
1989, the ASC listed only 687 SNe, and many of those were of
unknown type. By June, 2001 (updates are
available at \verb+http://merlino.pd.astro.it/~supern/+), the number of
events in the ASC had increased to 1910. Almost all of the SNe 
of the 1990's had been assigned types, and some of the pre-1990 data
had been improved. We therefore decided that an updated study
of the absolute-magnitude distributions would be timely.

It should be acknowledged that the characteristic uncertainty of the
apparent magnitudes listed in the ASC, perhaps 0.2 or 0.3 mag, is not
negligible for this purpose. Thanks to the tremendous interest in
using SNe~Ia as distance indicators for cosmology, studies of 
the SN~Ia absolute-magnitude distribution now
can be based on carefully selected samples of events for which both
the apparent magnitudes and the relative distances are known to high
accuracy \citep[e.g.][]{hamuy96}. The absolute-magnitude scatter in 
these samples can be further reduced in two ways. First, the host 
galaxy extinction can be taken into account for each SN using the SN's
$B-V$ color. Next, the fact that some SNe~Ia are intrinsically dimmer
than others can be accounted for by using a correction to the light-curve width such 
as the $\Delta m_{15}$ parameter \citep{phillips99} or the method of multicolor
light-curve shapes \citep[MLCS,][]{riess96}. With these corrections the
dispersion in $M_{B}$ for SNe~Ia can be reduced to 0.11 \citep{phillips99},  
making them extremely useful as distance indicators for cosmology. 

Our present study, which has little to add to our knowledge of the SN~Ia 
absolute-magnitude distribution, is directed more at the absolute-magnitude
distributions of the other SN types. As far as the other SN types are 
concerned, either the number of events for which
accurate peak apparent magnitudes have been reported remains small
(SNe~Ib, Ic, II-L, IIn) or the intrinsic dispersion in the peak
absolute magnitude is large (SNe~II-P). For these types a study such as
this one, based on all available data, can be useful.  For comparison
with the other types, SNe~Ia are included as well. 

\section{Data}

Most of the SN types and apparent magnitudes are taken from the ASC
(as of June, 2001).  Several changes in the spectral types are made
after examining spectra available to us.  Some additional data (mostly 
for recent SNe~Ia) are included from \cite{perl99}, 
\cite{riess99}, \cite{riess98}, \cite{leonard01}, \cite{cloc00} 
\cite{macri01}, \cite{math00}, \cite{vdbergh94}, 
\cite{iwamoto00} and \cite{hatano01}. 
Magnitudes from Perlmutter et al.\ include K-corrections.  
Several changes in the apparent magnitudes are made after
examining the Atlas of Light Curves of SNe~I by \cite{leib91}. 
Where possible we use blue magnitudes.
Photographic magnitudes are converted to blue
magnitudes using the usual relation $B=m_{pg}+0.3$. When necessary we
assume that $B=V=R$ (except for the K-corrected events). The apparent
magnitudes are corrected for Galactic extinction following \cite{sfd98}  
with the help of NED\footnote{The NASA/IPAC
Extragalactic Database (NED) is operated by the Jet Propulsion Laboratory,
California Institute of Technology, under contract with the National
Aeronautics and Space Administration.}.   

When possible, we use a Cepheid distance to the parent galaxy or to a
member of the same galaxy group as the parent galaxy.  The Cepheid
distances are obtained from \cite{freedman01}, \cite{macri01},
\cite{ferr00} and \citet{saha01a,saha01b}. 
If a Cepheid distance
is not available we use distances from the Nearby Galaxies Catalog
\citep[NGC;][]{tully88}, which incorporates a Virgocentric infall model to
relate galaxy recession velocity and distance. The errors in the
relative distances to these events contribute significantly to the
absolute-magnitude scatter in our figures. We rescale the distances
from the $H_{0}=75$ km~s$^{-1}$~Mpc$^{-1}$ of the NGC to our choice of
$H_{0}=60$.

If a Cepheid distance or an NGC distance is not available then the
luminosity distance is calculated \cite{ski00}, provided that 
the SN parent galaxy has a Galactocentric recession velocity of
$cz>2000$ km~s$^{-1}$. 
The luminosity distance is calculated assuming $H_{0}=60$,  
$\Omega_{m}=0.3$, and $\Omega_{\Lambda}=0.7$.  Errors in the relative
distances to these events should be small. A different choice of
$H_{0}$ would simply rescale the absolute magnitudes. Different
choices of $\Omega_{m}$ or $\Omega_{\Lambda}$ would change our plots
only slightly (at large distances). 
A few SNe whose parent galaxies have
$cz<2000$ km s$^{-1}$ but do not have a measured Cepheid distance and are 
not listed in the NGC are excluded from the study. The Large Magellanic
Cloud, the site of SN~1987A, is assigned a distance modulus of 18.50.

\section{Overview of Supernova Absolute Magnitudes}

In Figure~1, absolute magnitude is plotted against distance modulus
for all 297 SNe for which we have an estimate of the apparent
magnitude at the time of maximum light; we will refer to these as
\emph{maximum-light} SNe. Also shown, at the left side of Figure~1,
are estimates by \cite{schaefer96} of the absolute magnitudes of six
Galactic SNe of the last millennium that occurred within about
4~kpc of the Earth. Schaefer's estimates actually are of $M_{V}$,
rather than $M_{B}$, and his estimates of the distance moduli of these
events range from 11.4 to 12.7, off-scale to the left of Figure~1.
Figure~2 is like Figure~1, but for the 1078 SNe for which we
have only lower limits to the peak absolute brightness because the
apparent magnitude at the time of maximum light is unknown; we will
refer to these as \emph{limit} SNe. Both figures contain some straight
lines to guide the eye. The horizontal line at $M_{B}=-19.5$ is
regarded as characteristic of normal SNe~Ia and we will refer to this
as the SN~Ia \emph{ridge line}. The horizontal line at $M_{B}=-15$ and the
vertical lines at $\mu = 30$ and $\mu = 35$ will be referred to below.
The slanted lines correspond to apparent magnitudes of $B=16$ and
$B=25$. Figure~1 shows that so far most of the maximum-light
SNe have had peak apparent magnitudes brighter than $B=16$,
while Figure~2 shows that most of the limit SNe were fainter
than $B=16$ at their brightest observation. An indication of the
dramatic observational progress that has taken place during the 1990s
is that MB90 also plotted the $B=16$ line, and only \emph{four}
maximum-light SNe were to the right of it. MB90 did not
consider the limit SNe.

When viewing Figures~1 and 2 it is helpful to think in terms of
several different distance regimes. First come the Galactic
SNe, then SN~1987A and SN~1885A in the Local Group. Next come
SNe in ``nearby galaxies'', say within 10 Mpc ($\mu \leq
30$). The interval from $\mu=30$ to $\mu=35$ ($cz = 6000$ km\
s$^{-1}$, 100~Mpc) includes many SNe of the Local Supercluster,
with those of the Virgo cluster being concentrated near
$\mu=31.2$. From $\mu=35$ (or less) to $\mu\simeq40$ ($cz\simeq
60,000$ km s$^{-1}$, $z\simeq0.2$) is the ``Hubble flow'', where
luminosity distance depends on $H_{0}$ but only slightly on
$\Omega_{m}$ and $\Omega_{\Lambda}$. As will be seen below, most of
the maximum-light events in the Hubble flow are SNe~Ia;
this is due to the observational bias in favor of luminous
events. Beyond $\mu\simeq40$ is the realm of the ``high-redshift''
SNe where, fortunately for cosmology, the luminosity distance
depends significantly on $\Omega_{m}$ and $\Omega_{\Lambda}$. Here too, 
most of the events are SNe~Ia. At present most of the high-redshift events
appear in Figure~2, as limit SNe, but some of these will become
maximum-light SNe when their peak apparent magnitudes are
eventually reported.

\subsection{Subluminous Supernovae}

SN~1987A stimulated interest in subluminous SNe, e.g., those
having $M_{B} \geq -15$. The severe observational bias against them is
obvious from Figures~1 and 2. Beyond $\mu=30$, \emph{none} of the
maximum-light SNe are dimmer than $M_{B}=-15$.
\cite{schaefer96} considered two small volume-limited samples of
SNe and concluded that a substantial fraction of all SNe
are subluminous. One of his samples consisted of the six Galactic
SNe that are represented in Figure~1. Two of the six are
estimated to have been even less luminous than SN~1987A. Adopting
these estimates, \cite{hatano97} carried out a Monte Carlo simulation
of the visibility of Galactic SNe and concluded that it is quite likely 
that a significant fraction of the SNe in the Galaxy are ``ultra-dim''
($M_{V} > -13$).

Schaefer's other volume-limited sample consisted of SNe in
nearby galaxies.  In addition to SN~1987A there are four events in
Figure~1 that have $M_{B} > -15$.  SN~1940E (of unknown type) appeared
in the dusty, nearly edge-on galaxy NGC~253 and probably was heavily
extinguished so it may not have been intrinsically subluminous
\citep{schaefer96}.  SNe~1923A and 1945B, both in the nearly face-on
galaxy M83, do not appear to have been heavily extinguished \citep{schaefer96}. 
SN~1973R, in NGC~3727, was observed to be quite blue
\citep{patat93} and therefore also is unlikely to have been heavily
extinguished. It is likely that SNe~1923A (classified II-P only on the
basis of its light curve), 1945B (unknown type), and 1973R (Type II-P)
were intrinsically subluminous.

In Figure~1, among the Galactic SNe and the SNe in
nearby galaxies having $\mu\leq30$, seven of the
31 events are estimated to have had $M_{B}>-15$. Even this sample is
by no means free of bias against subluminous events, so it appears
that the fraction of all SNe that are subluminous is more than
(possibly much more than) one fifth. But because the number of such
events that have been seen is still so small, this fraction remains
very uncertain.

\subsection{Overluminous Supernovae}

The data on maximum-light SNe that were available to MB90
showed no convincing evidence for events that were significantly
overluminous with respect to the SN~Ia ridge line, but by now it has
become clear that overluminous events do exist, outstanding examples
being the Type IIn SN~1997cy \citep{germany00,turatto00}  
at $M_B \leq -20.30$ and the peculiar Type Ic SN~1999as \citep{hatano01} 
at $M_B \leq -21.60$. 

In Figure~1, the two brightest SNe are well above the SN~Ia 
ridge line, but the apparent magnitudes (and the types) of both are
uncertain. The light curves of SNe~1955B \citep[from][]{zwicky56} and  
1968A \citep[mainly][]{kaho68} are plotted and
compared to a template light curve for SNe~Ia by \cite{leib91}.
Whether or not these two events were overluminous depends on the
accuracy of the photographic photometry.  It is not clear that either 
of these events were genuinely overluminous.

The peak apparent magnitudes for SN~1920A and SN~1963S given in the
ASC would lead to $M_V= -22.2$ for both. However, the peak magnitudes
were obtained by extrapolating well beyond the observations using the
SN~Ia template light curve \citep{leib91}. Since both events 
are of unknown type, we treat them as limit SNe and use the
observations without extrapolation to estimate the brightest apparent
magnitudes.  For SN~1920A we use $m_{pg}=12.4$ (B. E. Schaefer 1999,
private communication)  
and obtain $M_B=-20.40$, still overluminous. For SN~1963S we
use $m_{pg}=15.00$ \citep{haro64} and obtain $M_B=-20.97$, also overluminous. 
 
In Figure~2, quite a few of the limit SNe have $M_B < -20$, but
undoubtedly this is mainly because limit SNe were not as well
observed as maximum-light SNe, so their magnitude errors tend
to be larger. In particular, many of the apparent magnitudes of the
high-redshift limit SNe are just first estimates that were made
at the time of discovery; if and when more accurate apparent
magnitudes are reported after the parent-galaxy light has been
subtracted, most of them will come down to the SN~Ia ridge line
(because most of them are SNe~Ia). The apparent magnitudes of nine of the 
most extreme cases in Figure~2, those that have $M_B \leq -21$ ---
SN~1999bd and 1961J (Type~II); 1988O (Type~Ia or Ic; see \S4.1); 1984M (unknown 
type); 1995av and 2000ei (probable Type II); 2000eh and 1999fo (Type~Ia); 
and SN~1971R (unknown type) --- 
are the preliminary estimates reported in the IAU Circulars.  We cannot be
sure that any of these events really were overluminous, but it would
take large magnitude errors to get them below $M_B=-20$.
 
At least two events in Figure~2, 1997cy and 
1999as, were genuinely overluminous.  Most of the others that appear to be 
overluminous in Figures~1 and 2 probably are spurious.  In Figure~1,
the fraction of the maximum-light events that have $M_B \leq -20$ is
0.067 (20 of 297).  In Figure~2, the fraction is 0.101 (109 of 1078).  
In view of the luminosity bias and the tendency of apparent-magnitude
errors to make ridge-line events appear to be overluminous, even the
0.067 is a generous upper limit to the true fraction of overluminous
SNe.  The true fraction probably is much lower than 0.01.

\section{Absolute-Magnitude Distributions by Supernova Type}

For the absolute-magnitude distributions we will consider only
SNe that have $\mu<40$ ($D<1\ Gpc$, $cz<60,000$ km s$^{-1}$). 
This restriction is imposed because as Figure~1 shows,
beyond $\mu=40$ the luminosity bias is so severe that few events are
much below the SN~Ia ridge line. We need to keep in mind that even the
distance-limited sample is strongly affected by luminosity bias ---
but to the extent that the bias is independent of the SN type,
we can obtain some useful comparative results.

Histograms that appear in the following figures refer to absolute
magnitudes that have been corrected only for Galactic extinction, not
for extinction in the parent galaxy. In addition to considering the
distributions of such uncorrected absolute magnitudes, we will also
consider ``intrinsic'' distributions, obtained as follows.
\cite{hatano98} used a Monte Carlo technique and a simple model of
the spatial distributions of SNe and dust in a characteristic
SN-producing galaxy to calculate extinction distributions for
each SN type, averaged over all galaxy inclinations.  Here, for
each SN type, we assume for simplicity a Gaussian intrinsic
absolute-magnitude distribution, convolve it with the appropriate
extinction distribution from Hatano et al., and vary the mean absolute
magnitude and dispersion of the intrinsic distribution to obtain the
best fit to the uncorrected distribution, as determined by the
Kolmogorov-Smirnoff (K-S) test \citep{press96}. Because very few
highly extinguished SNe make it into observational samples, we
use the ``extinction-limited subsets'' of Hatano et al., which exclude
extinctions larger than $A_{B}=0.6$ mag. (Changing this value by $\pm0.1$ mag
has very little effect on our results.) This simple statistical 
procedure makes a rough allowance for the effects of extinction on the
absolute-magnitude distributions.

\subsection{Type Ia}

SNe~Ia are plotted in Figure~3. Here, because of the relatively large
amount of data available, SNe~Ia whose apparent magnitudes were
designated as uncertain in the ASC are treated as limit
SNe. (For the other types discussed below, the number of events
is smaller and/or the absolute-magnitude dispersion is large, so we
accept those having uncertain apparent magnitudes as maximum-light
SNe.)

In Figure~3, the maximum light SNe~Ia do show a fairly high degree of
concentration to the ridge line at $M_{B}=-19.5$, as expected. The
scatter among the maximum-light events is due to a combination of
intrinsically subluminous SNe~1991bg-like events, extinction in the
parent galaxies, and errors in the apparent magnitudes and relative
distances.

Some of the most extreme SN~Ia are labeled in Figure~3. 
As mentioned above, the apparent magnitude of
the very bright limit events SN~1988O, 1999fo and 2000eh are uncertain. 
Also, although 1988O is listed as a Type~Ia in the ASC, from the published 
spectra \citep{stat94} it is hard to exclude Type~Ic. The five 
low--luminosity events that are labeled in Figure~3 include SNe
1991bg and 1957A, both of which are known to have been
spectroscopically peculiar and intrinsically subluminous.  SN~1986G is
known to have been both spectroscopically peculiar and highly
extinguished by dust in its parent galaxy. SNe~1996ai \citep{riess99}
and 1999cl \citep{kris00} are thought to have been highly extinguished
in their parent galaxies.  

The histogram in Figure~4 shows the uncorrected $M_B$ 
distribution for the 111 spectroscopically normal maximum--light SNe~Ia
that have $\mu \leq 40$. (If the spectroscopically low-luminosity
events were included the distribution could not be represented by a
gaussian.) Note that the distribution falls off steeply on the bright
side, as expected if SNe~Ia are thermonuclear disruptions of
Chandrasekhar-mass white dwarfs (or mergers of two white dwarfs)
because Type Ia light curves are powered by the radioactive decay of
$^{56}$Ni and its daughter $^{56}$Co and there is a hard upper limit
to the amount of $^{56}$Ni that one (or two) exploding white dwarfs
can eject. (This is a reason to suspect that SN~1988O was a Type~Ic
rather than a Type~Ia.) The uncorrected distribution has a mean
absolute magnitude of $\overline{M}_{B}=-19.16\pm0.07$, with a
dispersion about the mean of $\sigma=0.76$. The solid curve, the
gaussian intrinsic distribution obtained as described above, has
$\overline{M}_{B}=-19.46,\ \sigma=0.56$. The extinction correction has
made the mean absolute magnitude brighter by 0.3 mag., and produced a
moderate reduction in the dispersion. The dashed curve is the
convolution of the intrinsic distribution with the adopted extinction
distribution; this is to be compared with the uncorrected distribution
of the histogram. The K-S confidence level that the convolved
distribution is consistent with the uncorrected distribution is 89\%.

\subsection{Types Ib and Ic} 

Figures~5 and 6 are like Figures~3 and 4, but for SNe~Ibc (SNe~Ib and
Ic considered together because of the small amount of data available).
In Figure~5, five maximum light SNe~Ibc are significantly brighter
than the others.  The peculiar Type~Ic SN~1998bw, well known for its
possible connection to GRB980425, and the Type~Ib SN~1954A are well
observed.  The peculiar Type Ib SN~1991D is listed as a limit in the
ASC, but we have changed it to a maximum-light SN 
(S. Benetti et al., in preparation). For the Type~Ib SN~1992ar we have used 
the average of two apparent magnitudes estimated by \cite{cloc00}
using two different template light curves for SNe~Ic.  The apparent
magnitude of the Type~Ic SN~1999cq is from \cite{math00}.

The extremely bright limit event SN~1999as was discussed above.
The apparent magnitudes of the other other bright limits, SNe~1999bz
and 1993P (both Type~Ic), are uncertain.

The uncorrected distribution of the 18 maximum-light SNe~Ibc has
$\overline{M}_{B}=-17.92\pm0.30$, $\sigma=1.29$, and the intrinsic
distribution has $\overline{M}_{B}=-18.04$, $\sigma=1.39$. The K-S
confidence level is 96\% but the test tends to give
overly high confidence levels when the sample size is small.

Figure~5 raises the suspicion that there may be two separate
luminosity groups of SNe~Ibc, one perhaps tightly clustered near
$M_B\leq-19$ and another with a wider range at $M_B \simeq -17$. 
We will refer to these two groups as \emph{bright} and \emph{normal}. 
Because of this we have decided to consider a double peak distribution as
well as the single peak distribution. 
The double peak distribution is given by:
\begin{equation}
f(x)=f_0\biggl(w \exp\biggl[-\frac{(x-x_1)^2}{2\sigma_{1}^{2}}\biggr] +
\exp\biggl[-\frac{(x-x_2)^2}{2\sigma_{2}^{2}}\biggr]\biggr).  
\end{equation}
Here, $x_1$ and $\sigma_1$ are the mean absolute magnitude and dispersion, 
respectively, for the bright peak. The dim peak is similarly determined  
by $x_2$ and $\sigma_2$. The fifth parameter is the weighting factor, $w$.  
($f_0$ is the normalization constant.)
All five of these parameters are varied to get the best fit. 

Figure~7 is like Figure~6 but for a double peak distribution consisting 
of 5 bright and 13 normal SNe~Ibc. For the normal group, the 
uncorrected distribution has $\overline{M}_B = -17.23\pm0.17$, $\sigma=0.62$ 
and the intrinsic distribution has $\overline{M}_B = -17.61$, $\sigma=0.74$.  
For the bright group, the uncorrected distribution has 
$\overline{M}_B = -19.72\pm0.24$, $\sigma=0.54$ and the intrinsic distribution 
has $\overline{M}_B =-20.26$, $\sigma=0.33$. The weight parameter for
the intrinsic distribution is $w=0.28$.  
The K-S confidence level approaches unity for these small samples.  

The hydrogen-poor progenitors of SNe~Ibc have small radii so the light
curves of SNe~Ibc are powered by the radioactive decay of $^{56}$Ni.
Since the rise times to maximum light of SNe~Ibc and SNe~Ia are
similar, the ratio of the ejected masses of $^{56}$Ni in SNe~Ia and
SNe~Ibc is approximately given by the ratio of their peak
luminosities. Using $M_B =-19.46$ for SNe~Ia and $M_B =-18.04$ for 
SNe~Ibc would give a ratio of nickel masses of 3.70, e.g., if the
characteristic nickel mass of SNe~Ia is 0.6 $M_{\odot}$ then that of
SNe~Ibc would be 0.16 $M_{\odot}$. However, SNe~Ibc are redder than
SNe~Ia at maximum light and a smaller proportion of their luminosity
goes into the B band (cf. MB90), therefore the 0.16 $M_{\odot}$ may be
an underestimate of the average value. 
The nickel mass determined from the tail of the light curve for
SN~1994I was 0.07 $M_{\odot}$ \citep{young95}. Nickel
masses for SNe~Ibc may be strongly dependent on the nature of the
progenitor.

\subsection{Type II-L}

SNe~II-L are shown in Figures~8 and 9.  For the whole sample of 16
maximum-light SNe~II-L the uncorrected distribution has
$\overline{M}_B =-17.80\pm0.22$, $\sigma=0.88$ and the intrinsic
distribution has $\overline{M}_B =-18.03$, $\sigma=0.90$. The
K-S confidence level is 91\%. 

All four of the bright SNe~II-L --- SNe~1961F, 1979C, 1980K, and 1985L
--- were well observed.  As has been discussed by \cite{yb89},
\cite{gaskell92} and \cite{patat94} the available data suggest that it
may be appropriate to divide SNe~II-L into two luminosity groups;
which, again, we refer to as bright and normal. When we do this, with
only 4 events in the bright group and 12 in the normal group, we
obtain Figure~10.  For the normal group, the uncorrected distribution
has $\overline{M}_B =-17.36\pm0.12$, $\sigma=0.43$, and the intrinsic
distribution has $\overline{M}_B =-17.56$, $\sigma=0.38$.  For the
bright group, the uncorrected distribution has
$\overline{M}_B =-19.12\pm0.12$, $\sigma=0.23$, and the intrinsic
distribution has $\overline{M}_B =-19.27$, $\sigma=0.51$. The weight 
parameter for the intrinsic distribution is $w=0.43$. For these
small samples the confidence level, again, approaches unity. 

If, as is commonly assumed, the progenitors of SNe~II-L are red
supergiants that have lost much but not nearly all of their hydrogen
envelopes, then the peak luminosities of SNe~II-L depend primarily on
the radius of the progenitor star and in some cases perhaps also on
the strength of the circumstellar interaction, but not on the amount
of ejected $^{56}$Ni.  If this is correct, then the similarity of the
luminosities of the bright group of SNe~II-L and the luminosities of
normal SNe~Ia ($\overline{M}_B =-19.27$ and $-19.46$ respectively), as well 
as the similarities of the bright and normal SNe~II-L 
with the two possible groups of SNe~Ibc, must be coincidental. 

\subsection{Type II-P}

SNe~II-P are shown in Figures~11 and 12.  The uncorrected distribution of 
the 29 maximum-light SNe~II-P has $\overline{M}_{B}=-16.61\pm0.23$, 
$\sigma=1.23$. The intrinsic distribution has $\overline{M}_{B}=-17.00$,
$\sigma=1.12$ with a K-S confidence level approaching unity.  

The progenitors of SNe~II-P are red supergiants that have retained
large hydrogen masses. The large dispersion in the luminosities of
SNe~II-P is primarily due to a wide range in the radii of the
progenitor stars \citep{young94}.

\subsection{Type IIn}

Figures~13 and 14 show SNe~IIn.  The brightest maximum-light event,
SN~1983K, is listed as a Type~II-P in the ASC but we have treated it
as a Type~IIn in view of the narrow lines in its spectra \citep{niemela85}. 
None of the maximum-light SNe~IIn are extremely bright
or dim. However, among the limit SNe there are four that are
brighter than $M_{B}=-20$. SN~1997cy, discussed above, was genuinely
overluminous, while the apparent magnitudes of the other three
(SNe~1990S, 1995aa and 1999Z) are uncertain.

The uncorrected distribution of the 8 maximum-light SNe~IIn has
$\overline{M}_B = -18.78\pm0.31$, $\sigma = 0.92$, and the intrinsic
distribution has $\overline{M}_B = -19.15$, $\sigma = 0.92$. The K-S
confidence level approaches unity. SNe~IIn are on average more 
luminous than other SNe~II because of circumstellar interaction. 

\section{Summary}

At least 7 of 31 SNe in our Galaxy and in galaxies within 10
Mpc appear to have been subluminous ($M_B \geq -15$).  Considering
that even in this sample there is an observational bias against them,
it appears that more (perhaps much more) than one fifth of all
SNe are subluminous, but because the number of such events seen
so far is small, this fraction remains very uncertain.
 
Only 20 of 297 extragalactic maximum-light SNe appear to be
overluminous ($M_{B}<-20$). Considering the strong observational bias
in favor of them, and that observational errors produce spuriously
overluminous events, it is safe to conclude that the fraction of all
SNe that are overluminous must be much lower than 0.01.
 
The results of our comparative study of the absolute-magnitude
distributions of SNe in the distance-limited sample ($\mu<40$)
are summarized in Table~1.  The main differences between our results
and those of MB90 are that (1) it has become clear that overluminous
SNe ($M_B \leq -20$) exist; (2) the absolute-magnitude
dispersion of SNe~Ibc has increased due to the discovery of some
rather luminous events; and (3) we present results for SNe~IIn, which
on average are the most luminous type of core--collapse SNe.

Even though the SN discovery rate increased dramatically in the
1990's, the numbers of subluminous events, overluminous events,
SNe~II-L, SNe~Ibc, and SNe~IIn for which apparent magnitudes at
maximum light are available are still small --- this is especially so
if SNe~II-L and Ibc should be divided into two luminosity groups.
Except for SNe~Ia and II-P, the study of SN absolute-magnitude
distributions remains data starved.  Systematic programs of discovery
and observation of events in the Hubble flow, such as the Nearby Supernova
Factory \citep{alder01} underway at the Lawrence Berkeley National 
Laboratory, should substantially improve the situation during the coming
years.

\clearpage

\begin{figure}
\plotone{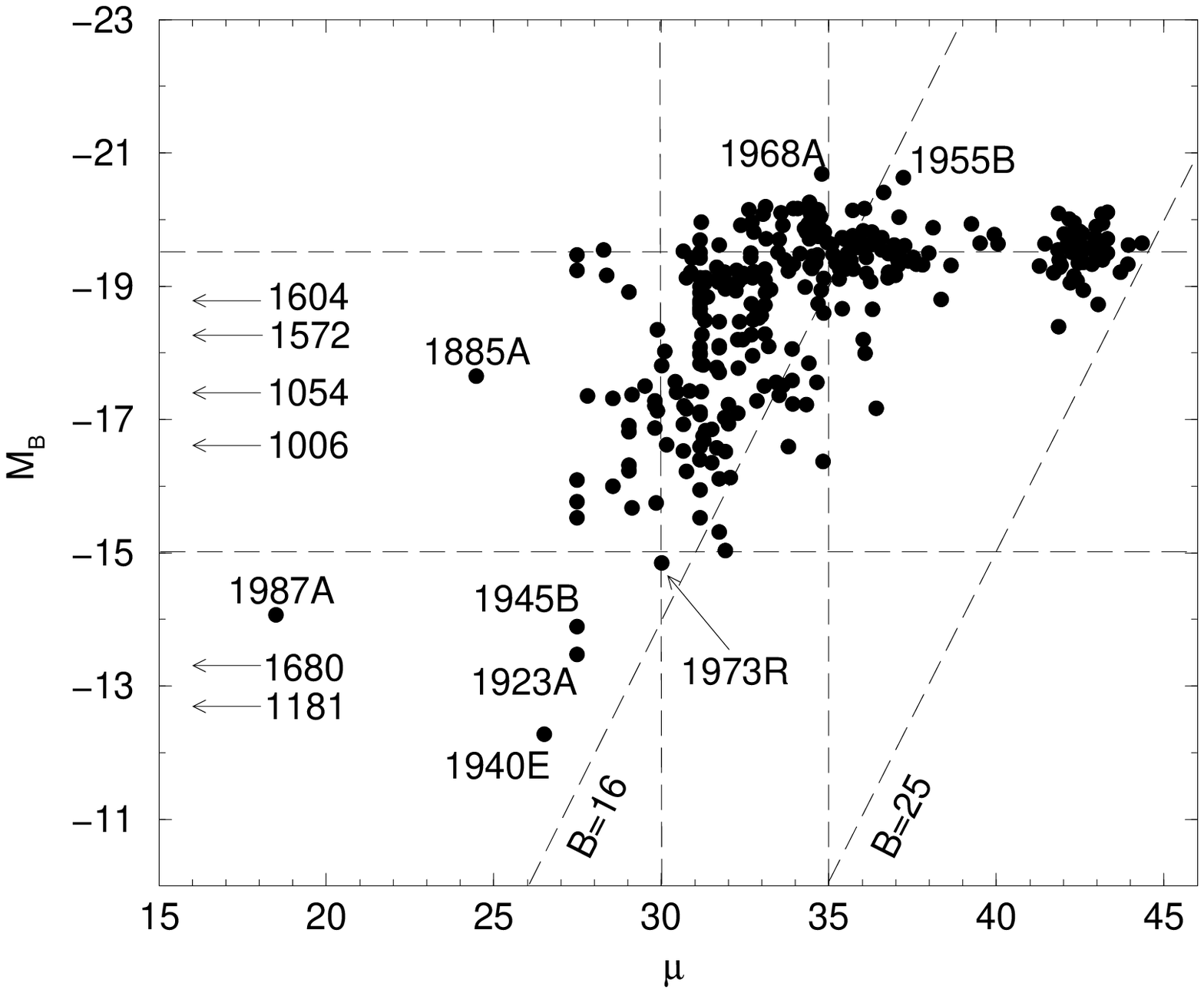}
\caption{\label{fig1} Absolute magnitude is plotted against distance 
modulus for all 297 maximum-light SNe, regardless of type.}  
\end{figure}

\begin{figure}
\plotone{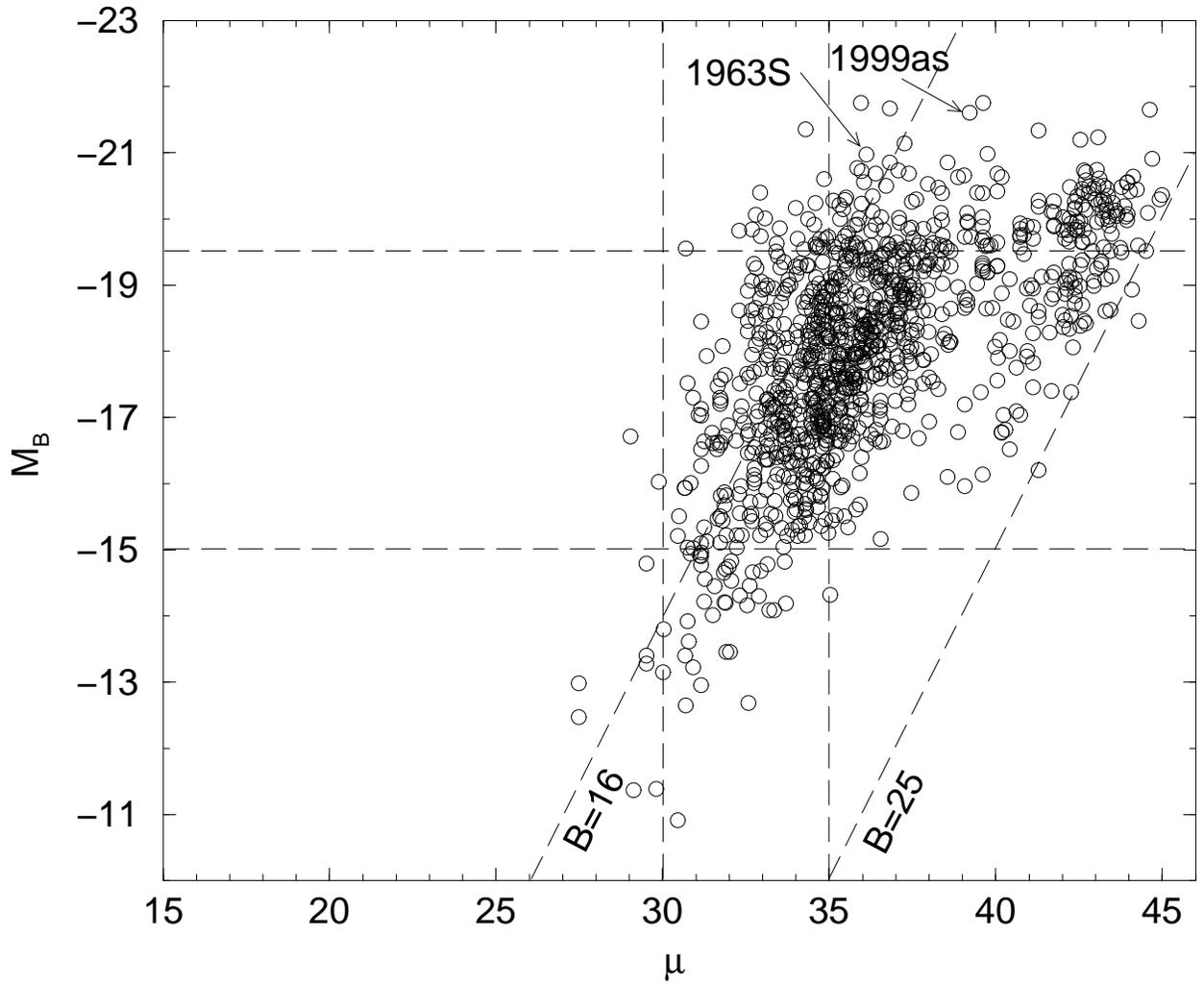}
\caption{\label{fig2} Absolute magnitude is plotted against distance modulus 
for all 1078 limit SNe, regardless of type.}
\end{figure}

\begin{figure}
\plotone{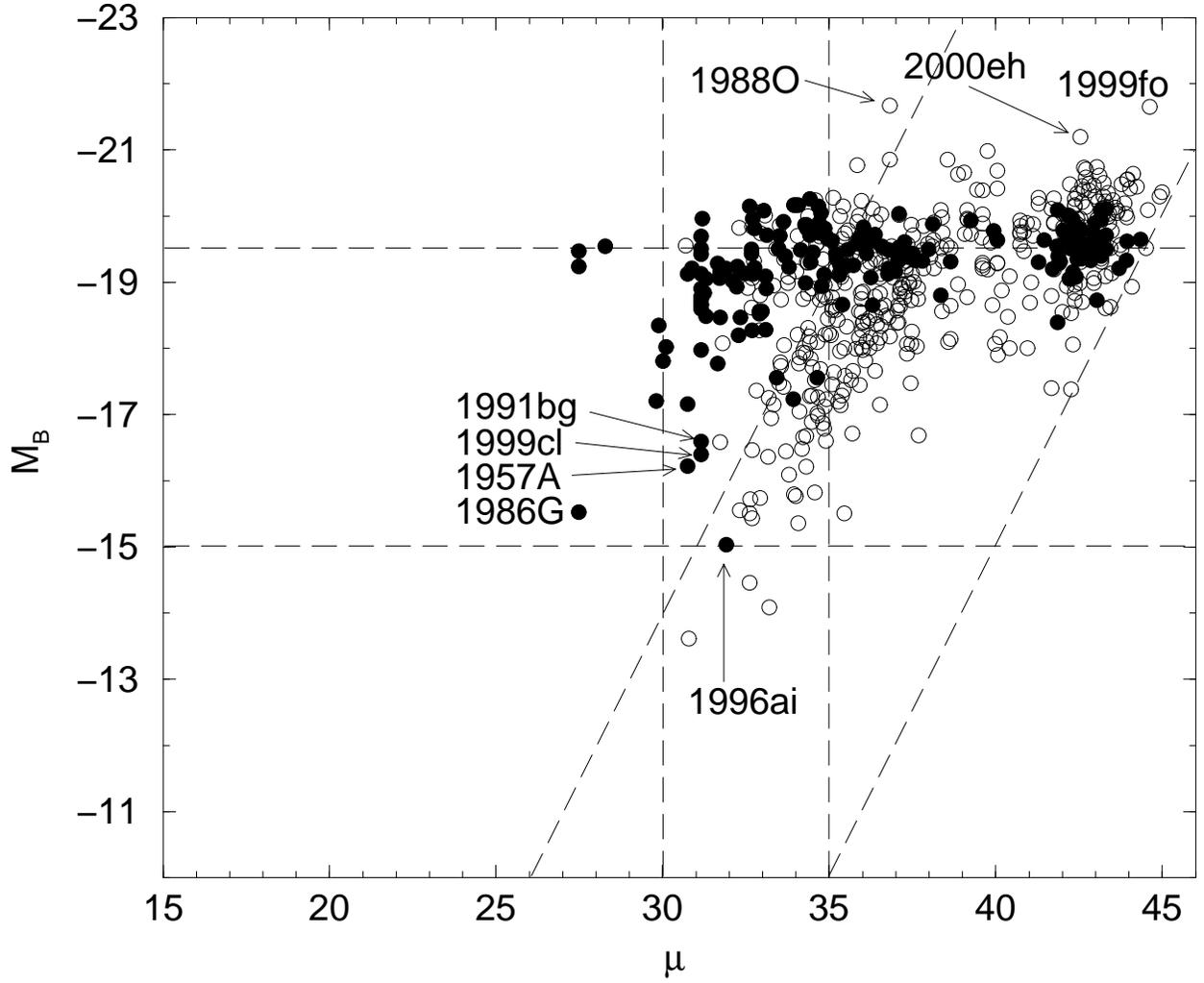}
\caption{\label{fig3} Absolute magnitude is plotted against distance modulus for 
SNe~Ia.  Filled symbols are maximum--light 
SNe (174) and open symbols are limit SNe (395).} 
\end{figure}

\begin{figure}
\plotone{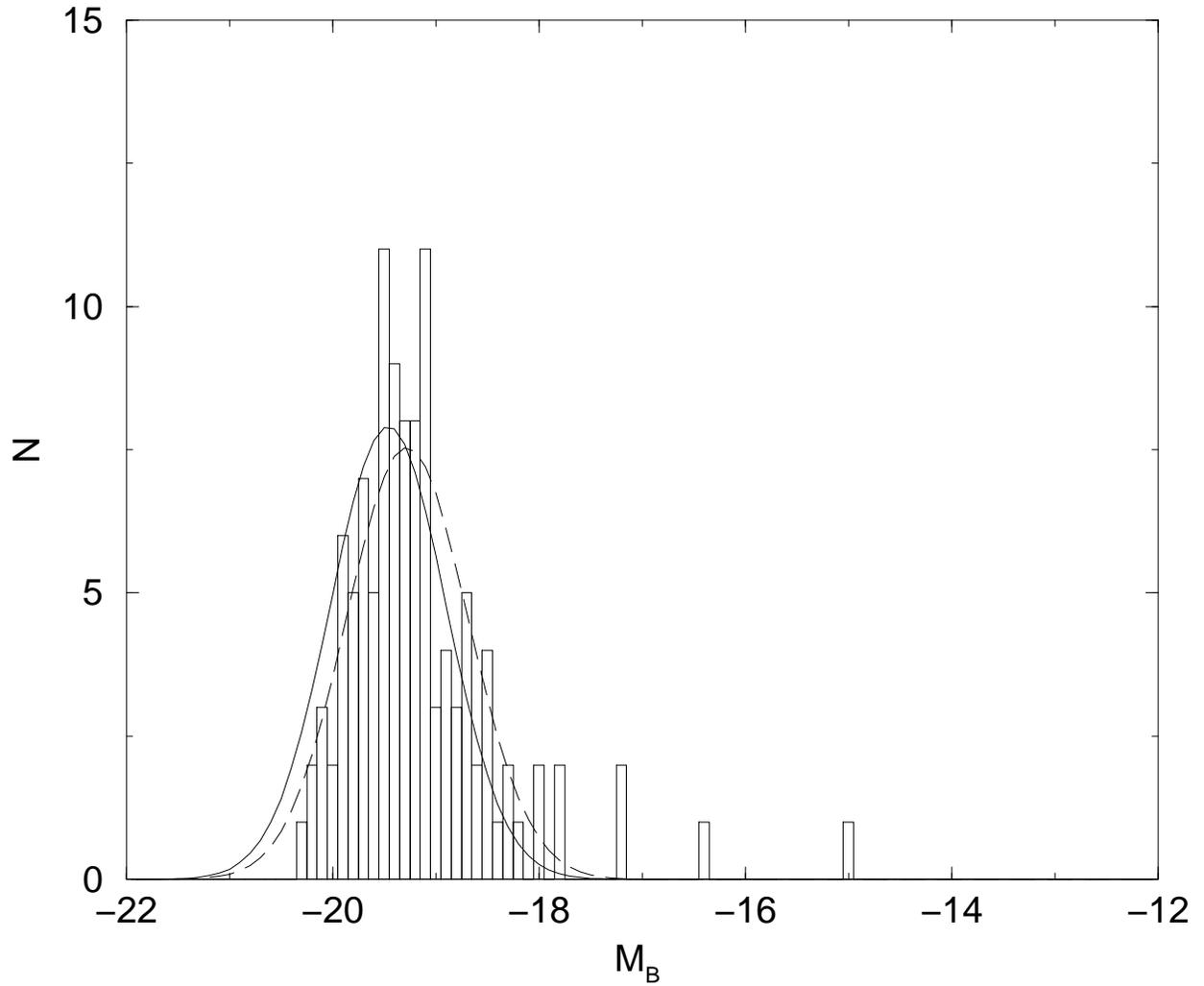}
\caption{\label{fig4} The absolute--magnitude distribution of 111 normal 
SNe~Ia at maximum-light 
having $\mu \leq 40$, uncorrected for extinction in the parent
galaxies (histogram) is compared with the convolution (dashed line) of
the adopted intrinsic gaussian distribution (solid line) with the
adopted parent-galaxy extinction distribution (see text).} 
\end{figure}

\begin{figure}
\plotone{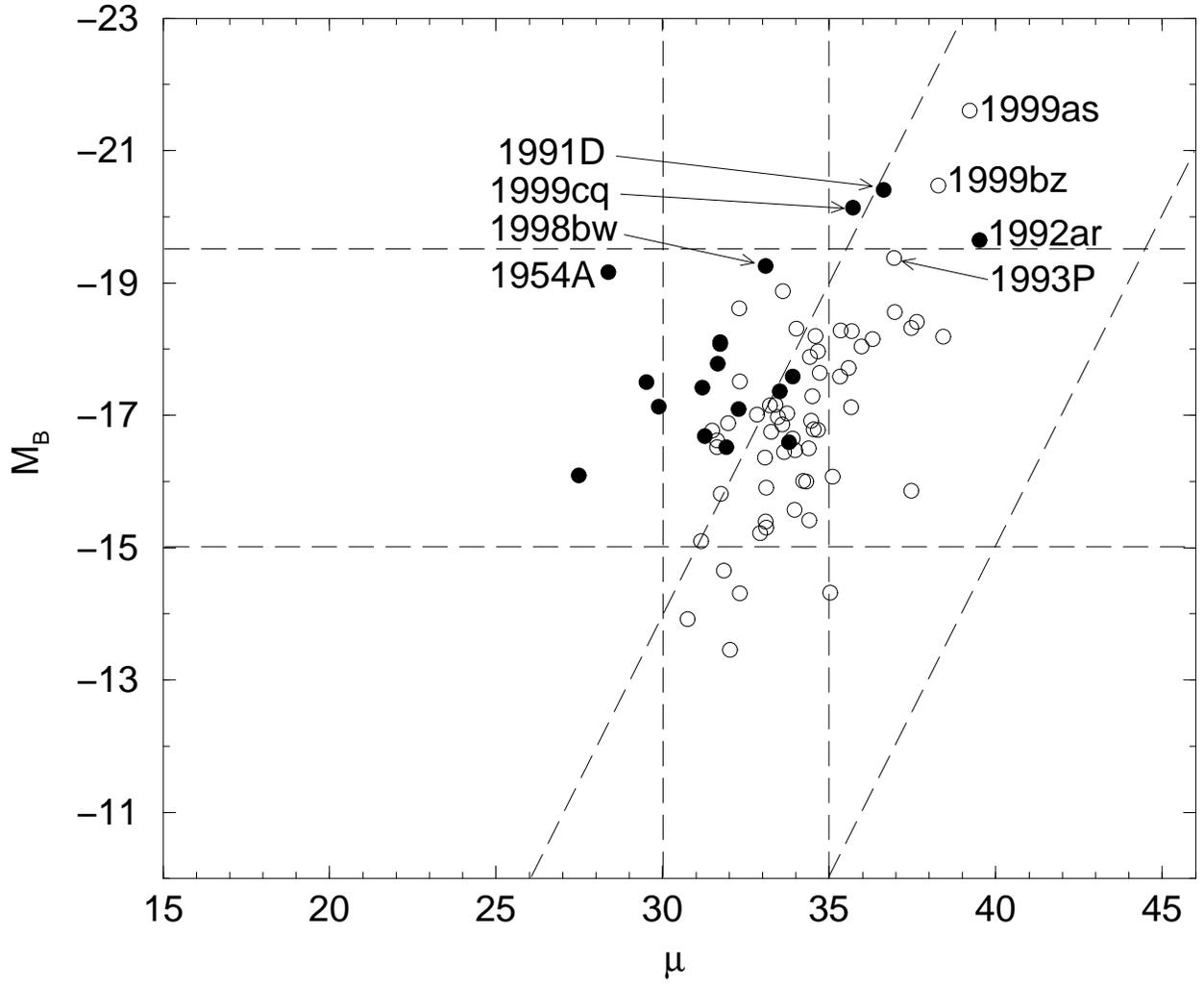}
\caption{\label{fig5} Like Figure~3 but for SNe~Ibc. (18 maximum-light
SNe and 59 limit SNe.) } 
\end{figure}

\begin{figure}
\plotone{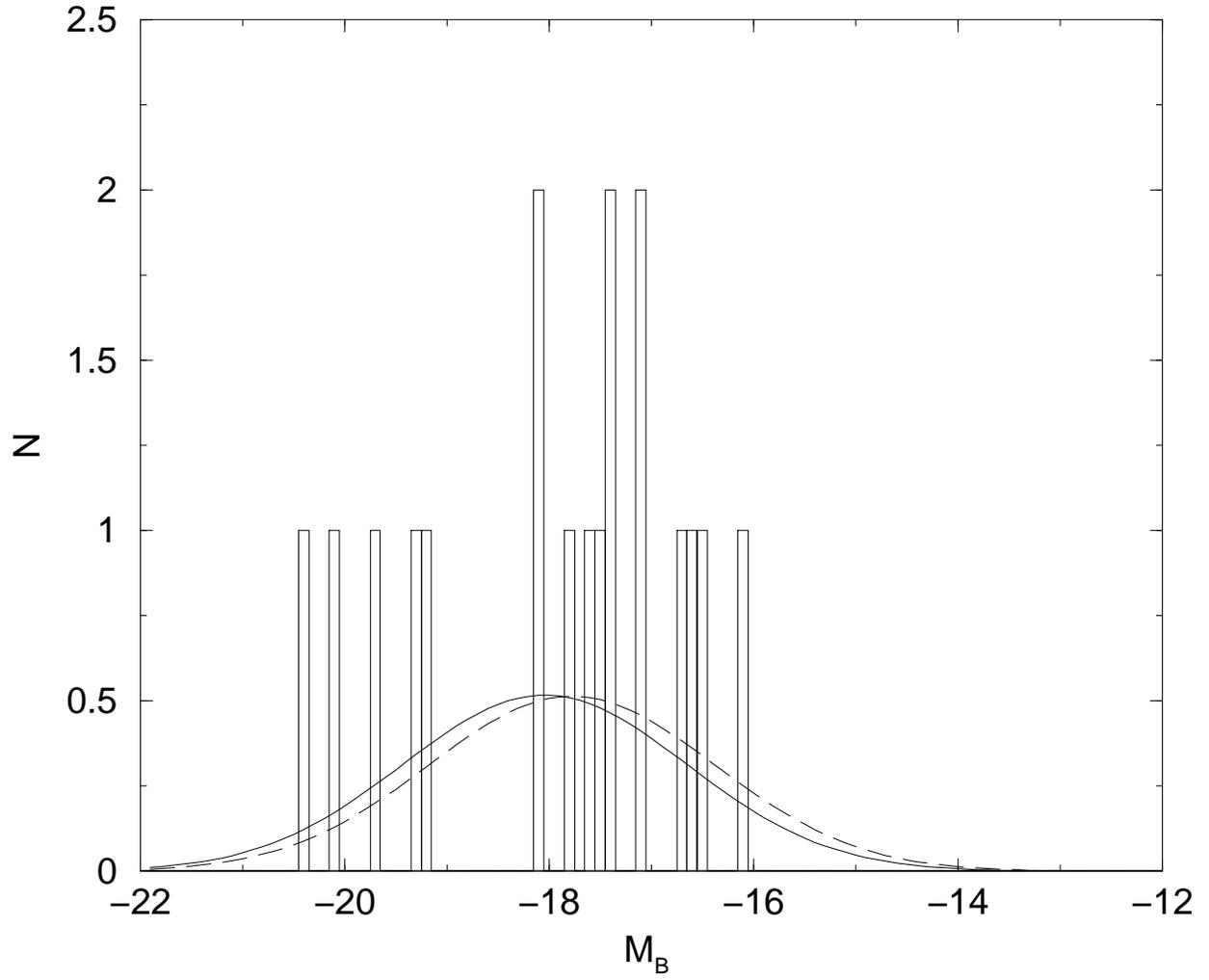}
\caption{\label{fig6} Like Figure~4 but for 18 SNe~Ibc.}
\end{figure}

\begin{figure}
\plotone{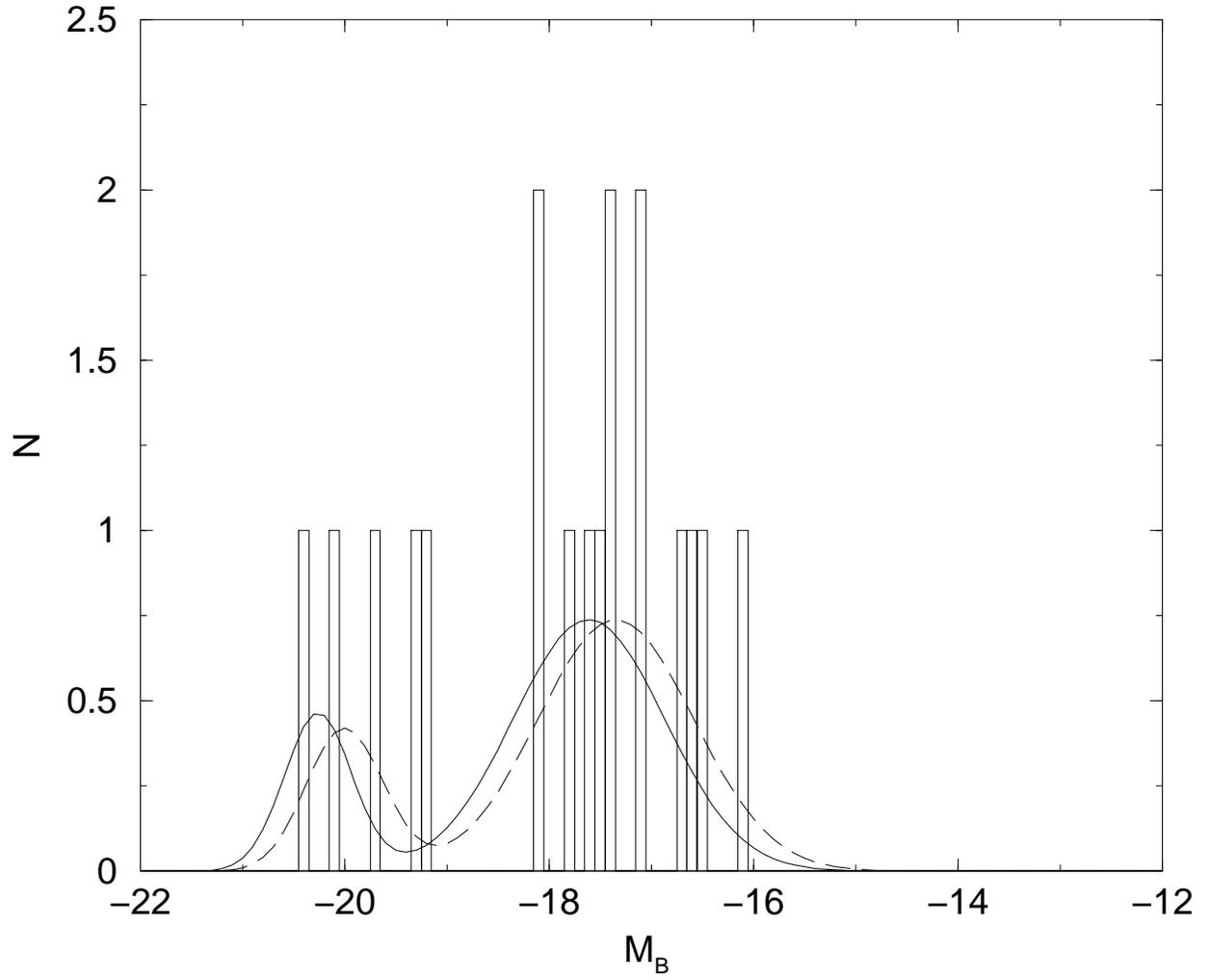}
\caption{\label{fig7} Like Figure~6 but for SNe~Ibc divided into two groups of 5
bright and 13 normal events. } 
\end{figure}

\begin{figure}
\plotone{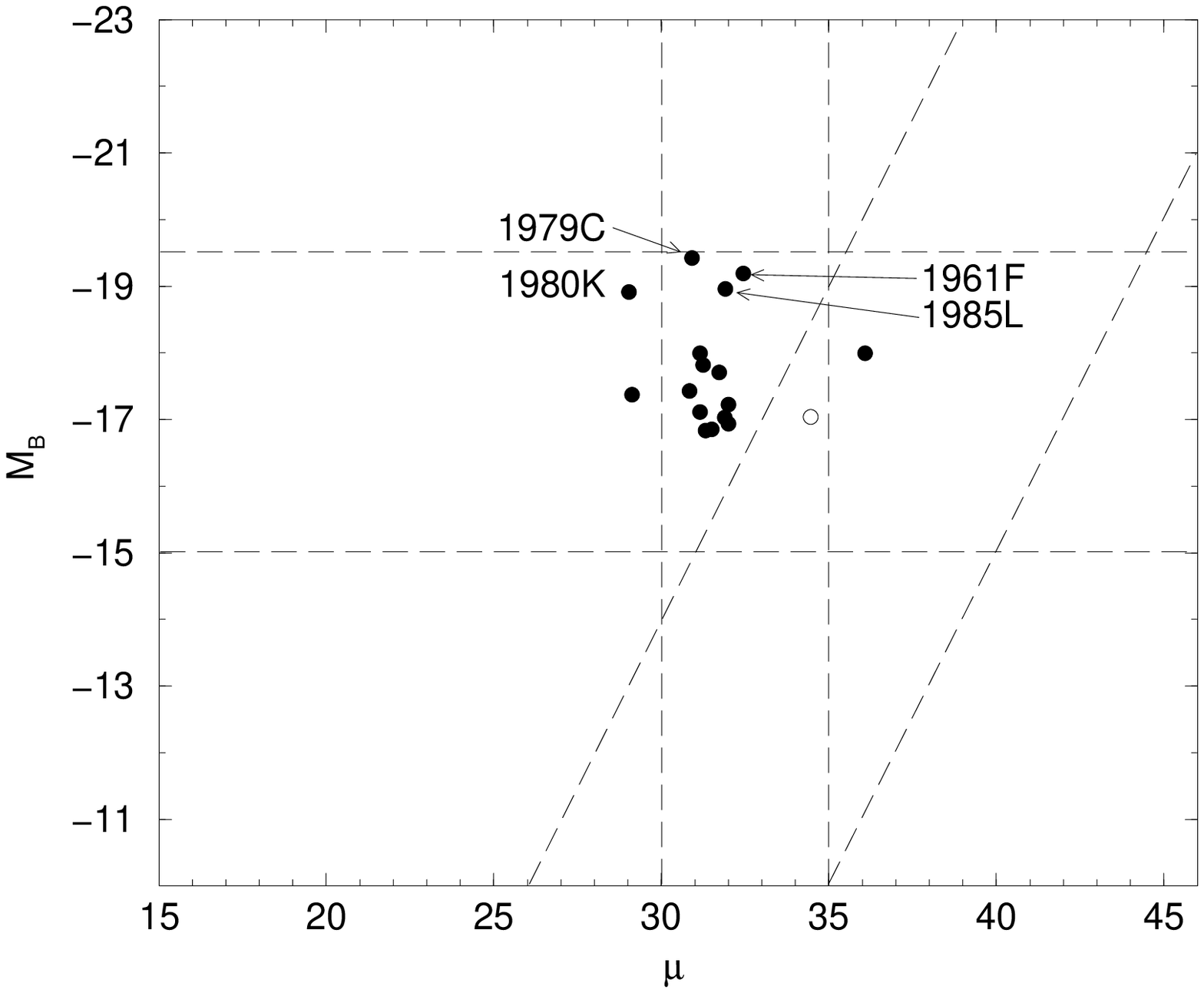}
\caption{\label{fig8} Like Figure~3 but for SNe~II-L. (16 maximum-light
SNe and 1 limit SN.) }
\end{figure}

\begin{figure}
\plotone{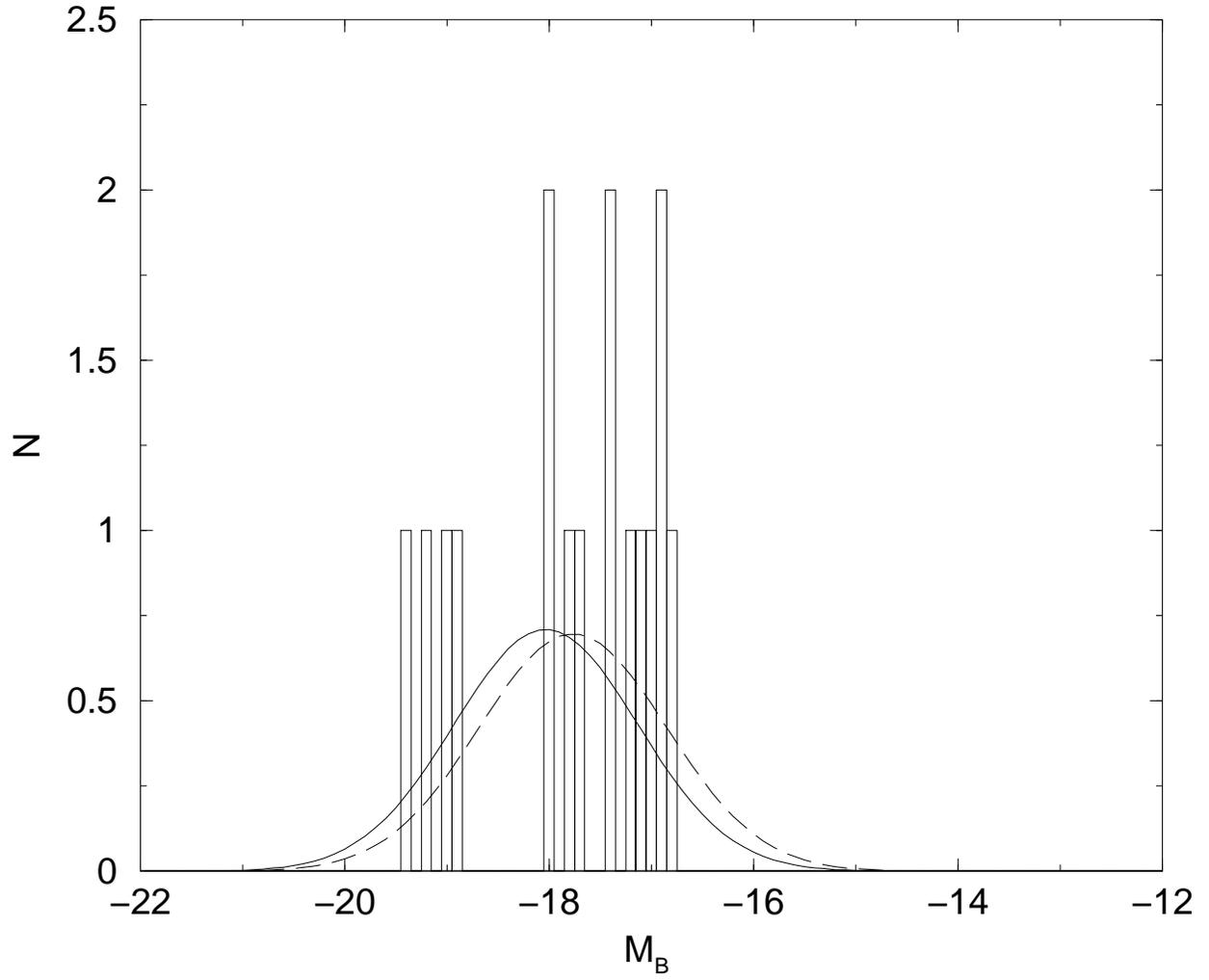}
\caption{\label{fig9} Like Figure~4 but for 16 SNe~II-L. }
\end{figure}

\begin{figure}
\plotone{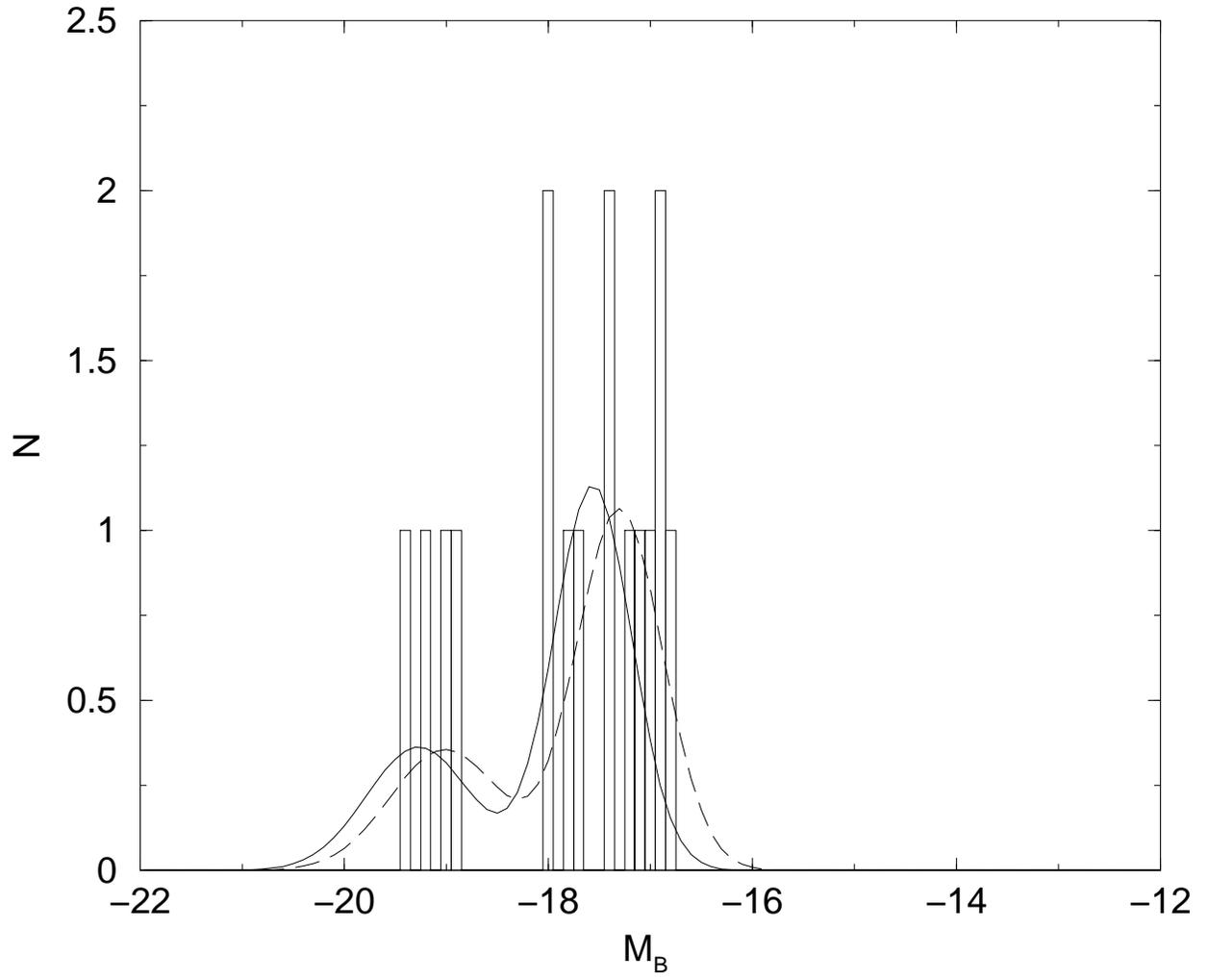}
\caption{\label{fig10} Like Figure~7 but for SNe~II-L. (4
bright and 12 normal events.) }
\end{figure}
 
\begin{figure}
\plotone{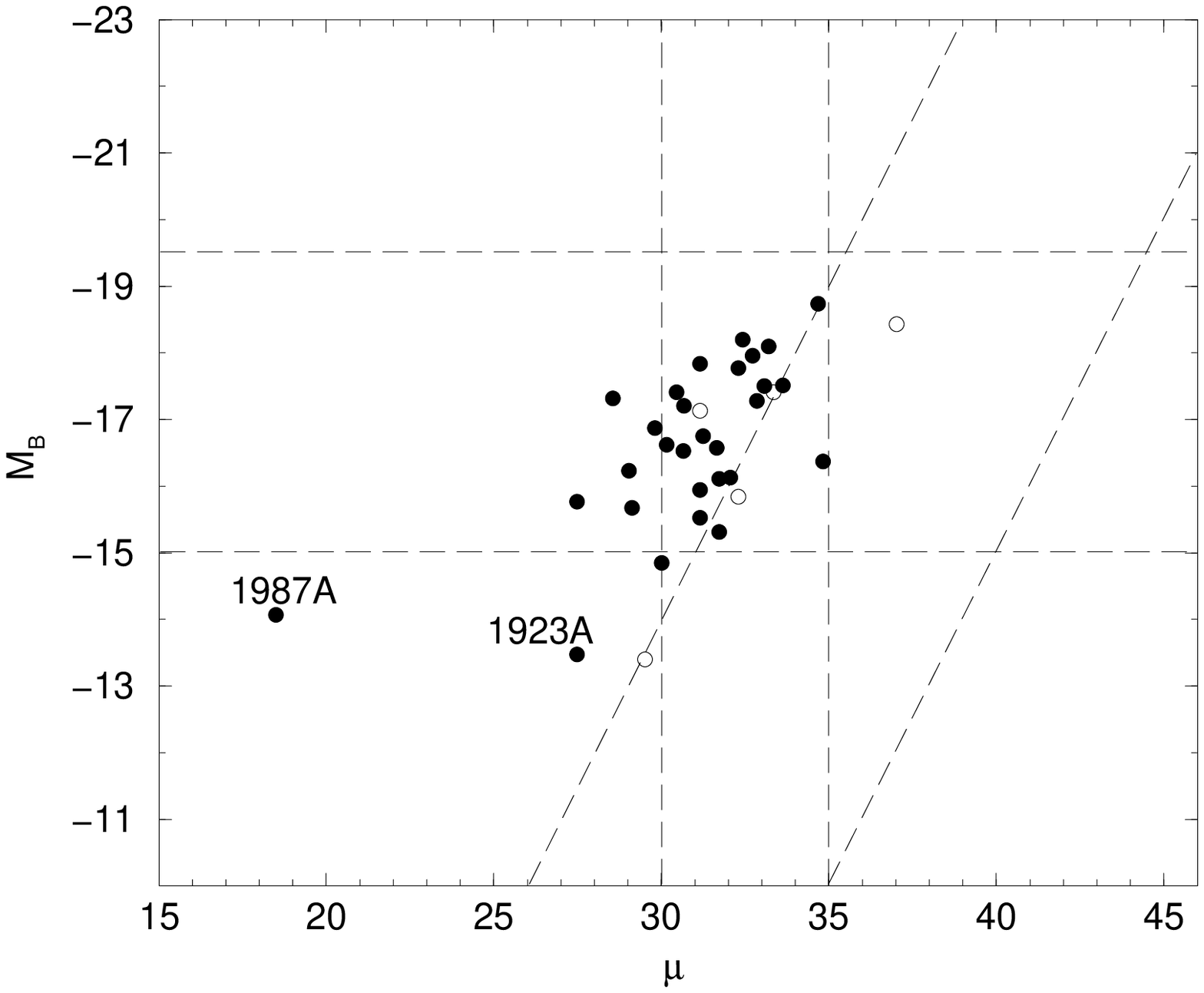}
\caption{\label{fig11} Like Figure~3 but for SNe~II-P. (29 maximum-light
SNe and 5 limit SNe.) }
\end{figure}

\begin{figure}
\plotone{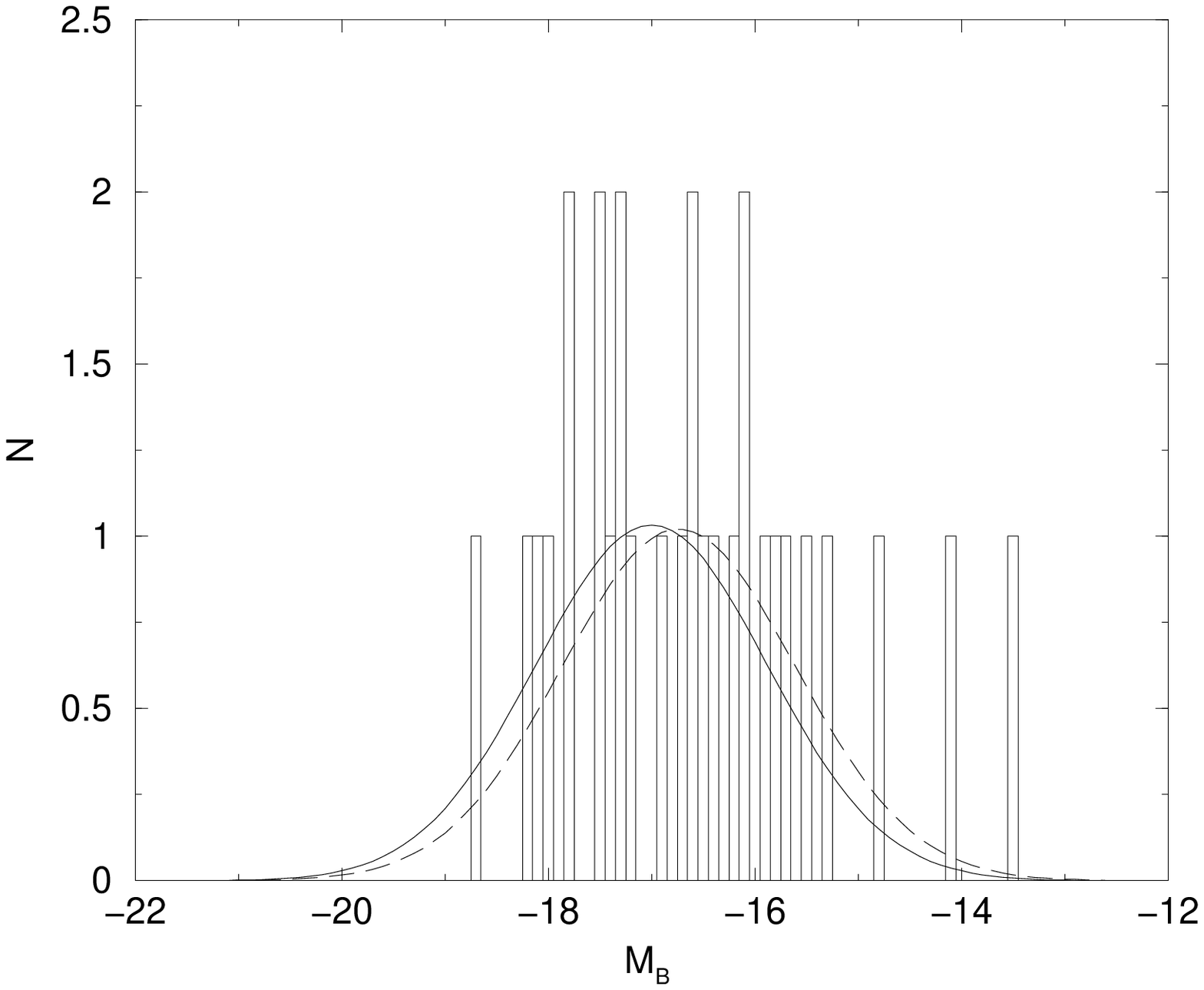}
\caption{\label{fig12} Like Figure~4 but for 29 SNe~II-P. }
\end{figure}

\begin{figure}
\plotone{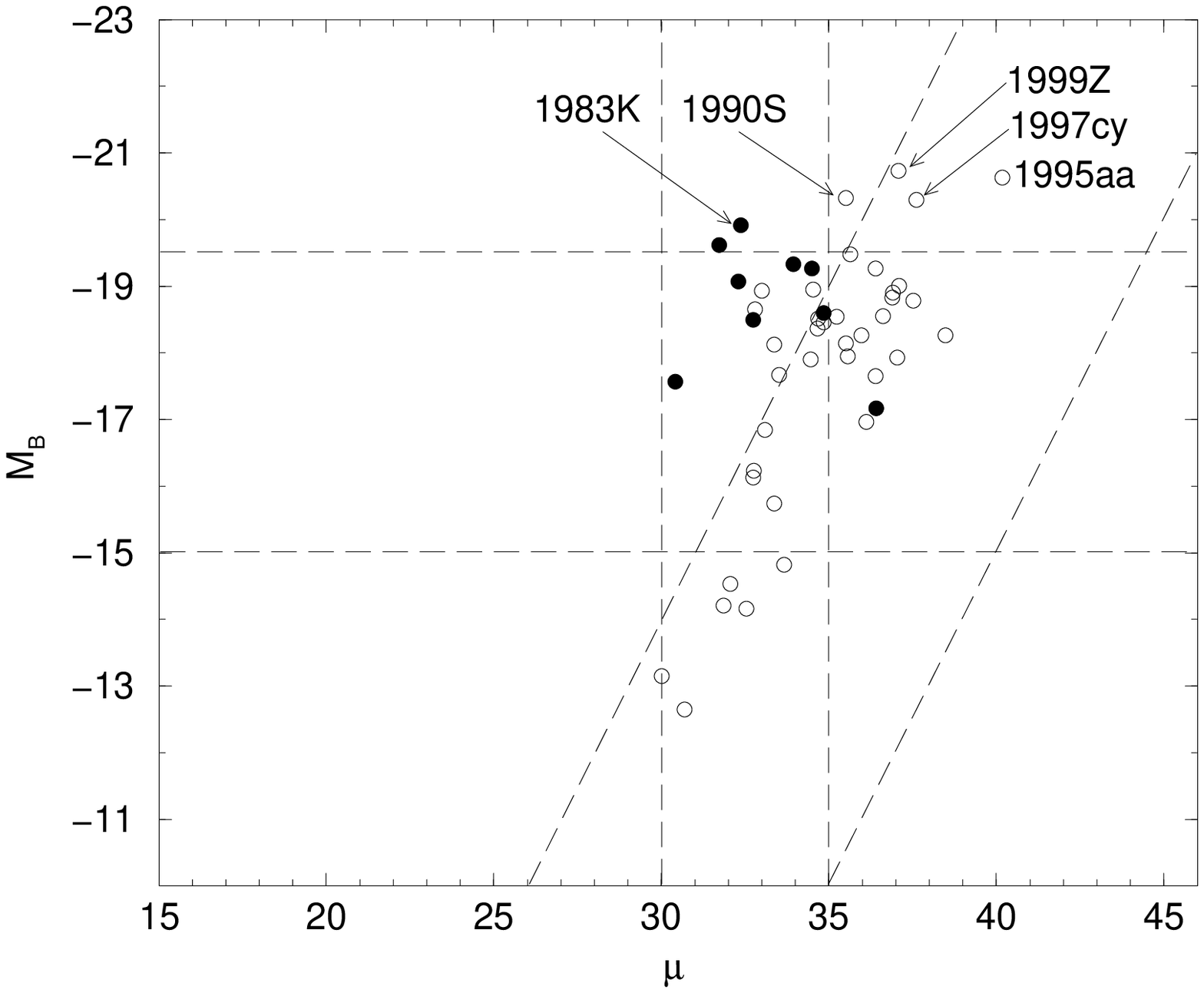}
\caption{\label{fig13} Like Figure~3 but for SNe~IIn. (9 maximum-light
SNe and 38 limit SNe.) }
\end{figure}
 
\begin{figure}
\plotone{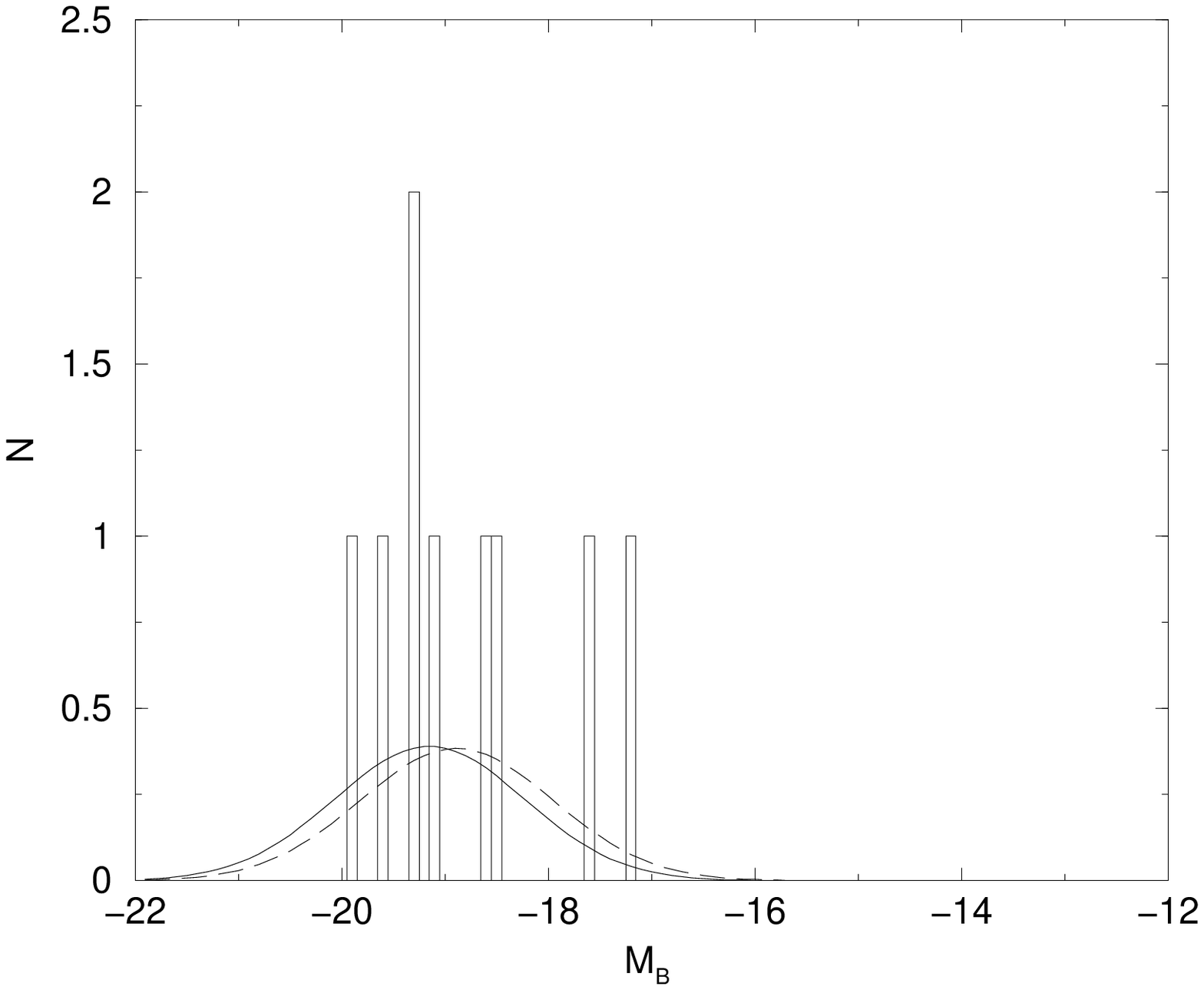}
\caption{\label{fig14} Like Figure~4 but for 9 SNe~IIn. }
\end{figure}

\clearpage
 
\begin{deluxetable}{lrrrrcc}
\footnotesize
\tablecaption{\label{table1} Results.}
\tablewidth{0pt}
\tablehead{
\colhead{SN Type} &
\colhead{$\overline{M}_{B,obs}$} &
\colhead{$\sigma_{obs}$} &
\colhead{$\overline{M}_{B,int}$} &
\colhead{$\sigma_{int}$} &
\colhead{Conf.}  &
\colhead{N}
}
\startdata
Normal Ia &$-19.16\pm 0.07$ &0.76 &$-19.46$ &0.56 &0.89 &111  \\
Total Ibc   &$-17.92\pm 0.30$ &1.29 &$-18.04$ &1.39 &0.96 &18 \\
Bright Ibc   &$-19.72\pm 0.24$ &0.54 &$-20.26$ &0.33 &$\sim1$ &5 \\
Normal Ibc   &$-17.23\pm 0.17$ &0.62 &$-17.61$ &0.74 &$\sim1$ &13 \\
Total II-L &$-17.80\pm 0.22$ &0.88 &$-18.03$ &0.90 &0.91 &16 \\
Bright II-L &$-19.12\pm 0.12$ &0.23 &$-19.27$ &0.51 &$\sim1$ &4  \\
Normal II-L &$-17.36\pm 0.12$ &0.43 &$-17.56$ &0.38 &$\sim1$ &12   \\
II-P   &$-16.61\pm 0.23$ &1.23 &$-17.00$ &1.12 &$\sim1$ &29  \\
IIn   &$-18.78\pm 0.31$ &0.92 &$-19.15$ &0.92 &$\sim1$ &9  \\
\enddata
\end{deluxetable}

\eject
\end{document}